\documentclass[prb,amsmath,superscriptaddress,showpacs,showkeys,twocolumn]{revtex4-1}
\usepackage{amssymb,amsmath}   % need for subequations
\usepackage[dvips]{graphicx}   % need for figures
\usepackage{verbatim}   % useful for program listings
\usepackage{color}      % use if color is used in text
\usepackage{subfigure}  % use for side-by-side figures
\usepackage{hyperref}   % use for hypertext links, including those to external documents and URLs
\usepackage{gensymb}
\usepackage{epstopdf}
\usepackage{natbib}
\usepackage{enumerate}

\begin{document}

\title{Rabi noise spectroscopy of individual two-level tunneling defects}

\author{Shlomi Matityahu}
\affiliation{Department of Physics, Ben-Gurion University of the Negev, Beer Sheva 84105, Israel}
\affiliation{Department of Physics, NRCN, P.O. Box 9001, Beer-Sheva 84190, Israel}
\author{J\"urgen Lisenfeld}
\affiliation{Physikalisches Institut, Karlsruhe Institute of Technology (KIT), 76131 Karlsruhe, Germany}
\author{Alexander Bilmes}
\affiliation{Physikalisches Institut, Karlsruhe Institute of Technology (KIT), 76131 Karlsruhe, Germany}
\author{Alexander Shnirman}
\affiliation{Institut f\"ur Theorie der Kondensierten Materie, KIT, 76131 Karlsruhe, Germany}
\author{Georg Weiss}
\affiliation{Physikalisches Institut, Karlsruhe Institute of Technology (KIT), 76131 Karlsruhe, Germany}
\author{Alexey V. Ustinov}
\affiliation{Physikalisches Institut, Karlsruhe Institute of Technology (KIT), 76131 Karlsruhe, Germany}
\affiliation{Russian Quantum Center, National University of Science and Technology MISIS, Leninsky prosp. 4, Moscow, 119049, Russia}
\author{Moshe Schechter}
\affiliation{Department of Physics, Ben-Gurion University of the Negev, Beer Sheva 84105, Israel}

\date{\today}
\begin{abstract}
Understanding the nature of two-level tunneling defects is important for minimizing their disruptive effects in various nano-devices. By exploiting the resonant coupling of these defects to a superconducting qubit, one can probe and coherently manipulate them individually. In this work we utilize a phase qubit to induce Rabi oscillations of single tunneling defects and measure their dephasing rates as a function of the defect's asymmetry energy, which is tuned by an applied strain. The dephasing rates scale quadratically with the external strain and are inversely proportional to the Rabi frequency. These results are analyzed and explained within a model of interacting standard defects, in which pure dephasing of coherent high-frequency (GHz) defects is caused by interaction with incoherent low-frequency thermally excited defects.
\end{abstract}

%\pacs{85.75.Hh, 75.76.+j, 72.25.Dc, 75.70.Tj}
 \maketitle
%
%\section{Introduction} \label{Introduction} 
Since the early 1970's, various experiments over a wide range of amorphous solids revealed a universality in their thermal, acoustic and dielectric properties below $1\,$K.~\cite{ZRC71,BJF88,PRO02} In an attempt to account for this universal behavior, the existence of two-level tunneling defects as a generic property in amorphous systems was postulated.\cite{PWA72,APW72} This phenomenological standard tunneling model (STM) explains many of the universal low-temperature properties of the amorphous state of matter.~\cite{PWA87} However, despite extensive efforts, the exact nature of the two-level systems (TLSs) remains unknown. %by Phillips~\cite{PWA72} and independently by Anderson, Halperin, and Varma.~\cite{APW72}

With the recent progress in fabrication, manipulation, and measurement of quantum devices it became crucial to understand the microscopic nature of the environment responsible for decoherence. There exists abundant experimental evidence that TLS-baths form an ubiquitous source of noise in various devices such as superconducting microwave resonators,~\cite{ZJ12} single-electron transistors,~\cite{PA14} and nanomechanical resonators.~\cite{AKH03} In superconducting qubits, TLSs residing in the amorphous tunnel barrier of Josephson junctions were found to constitute a major source of decoherence.~\cite{SRW04,CKB04} Via their electric dipole moment, TLSs couple to the ac microwave fields in the circuit.~\cite{CJH10} Whereas this coupling is deleterious from the point of view of qubit operation, it opens up the possibility to use superconducting qubits as tools for detection, manipulation and characterization of individual TLSs. The transfer of an arbitrary quantum state from a superconducting phase qubit to a resonant TLS was first demonstrated by Neeley {\it et al.}.~\cite{NM08} This method was used to probe the coherence times of individual TLSs.~\cite{NM08,SY10} Furthermore, in Ref.~\onlinecite{LJ10a} it was shown that there exists an effective qubit-mediated coupling between TLSs and an externally applied electromagnetic ac field. This effective coupling was utilized to directly control the quantum state of individual TLSs by coherent resonant microwave driving.~\cite{LJ10b}

Previously,~\cite{LJ16} we measured the Ramsey (free induction decay) and spin-echo pure dephasing rates of individual TLSs in a phase qubit as a function of their asymmetry energy, which was tuned by an applied strain via a piezo actuator.~\cite{GGJ12} Since the mutual longitudinal coupling between TLSs is proportional to the product of their asymmetry energies, strain-tuning allows one to gradually increase the longitudinal coupling of a single probed TLS to a bath of other TLSs and study its dephasing rates as a function of this coupling. This yields information about the spectrum of the environment to which a TLS couples, and provides a test to distinguish between different TLS models. The experimental data on Ramsey dephasing indicate that the main low-frequency noise is quasi-static and can be attributed to slow thermal TLSs (with energy splitting smaller than the temperature), which flip between their energy eigenstates with maximum rates $\Gamma_{1,\mathrm{max}}\approx 10\,$(ms)$^{-1}$,~\cite{LJ16,MS16} much smaller than the dephasing rates of the probed TLS which are of the order of $\Gamma_{\mathrm{Ramsey}}\approx 1\,$$(\mu$s)$^{-1}$. For such an environment, the echo protocol should be very efficient. Surprisingly, the experiment shows that the echo dephasing rates are not negligible, revealing the existence of an additional noise source with a flat power spectral density. It was suggested that this white noise may arise due to fast relaxing TLSs that interact much more strongly with strain fields compared to the weakly interacting TLSs of the STM,~\cite{MS16,SM13} or may be a result of quasiparticle excitations.~\cite{LJ16,ZS16,BA16}

Here we study experimentally and theoretically the decoherence of Rabi oscillations of individual TLSs as a function of their strain-tuned asymmetry energy. At resonance with the driving field, the Rabi decay rate consists of three contributions from noise at different frequencies.~\cite{HJ08} The first contribution is due to noise at frequency equal to the energy splitting of the probed TLS ($\approx 2\pi\cdot7\,$GHz), and arises from degrees of freedom other than thermal TLSs, such as phonons or microwave photons. The other two contributions are due to noise at the Rabi frequency of the probed TLS (several MHz) and low-frequency quasi-static noise similar to the one responsible for the Ramsey dephasing discussed in Refs.~\onlinecite{LJ16} and~\onlinecite{MS16}. The last two contributions result from a transverse noise in the rotating frame of reference, the origin of which is suggested to be thermal TLSs. Due to its transverse nature, this noise leads to a quadratic strain dependence of the Rabi dephasing rate near the symmetry point of the probed TLS, making the dephasing rates smaller than those in a Ramsey experiment.

We begin with a model of a high-frequency (GHz) single TLS driven by a microwave field at frequency $\omega_{d}$ and interacting with a thermal bath. In the basis of local states of the TLS, the Hamiltonian of the system is (we set $\hbar=k_{B}=1$)
\begin{align}
\label{eq:Hamiltonian1}\hat{\mathcal{H}}=&\,\frac{1}{2}\left(\varepsilon\hat{\tau}_{z}+\Delta\hat{\tau}_{x}\right)-\Omega^{0}_{\mathrm{R}}\cos\left(\omega_{d}t\right)\hat{\tau}_{z}+\frac{1}{2}\hat{\tau}_{z}\hat{O}+\hat{\mathcal{H}}_{\mathrm{b}}\, .
\end{align}
The first term describes the TLS, characterized by the asymmetry and tunneling energies, $\varepsilon$ and $\Delta$, with $\hat{\tau}_{x}$ and $\hat{\tau}_{z}$ being the Pauli matrices. The second term describes its coupling to the driving field with $\Omega^{0}_{\mathrm{R}}=pE_{z}$ being the maximum Rabi frequency, where $p$ is the electric dipole moment of the TLS and $E_{z}$ is the component of the electric field along its dipole moment. The third term is the coupling of the TLS to the bath observable $\hat{O}$. Motivated by the experimental data below, we write $\hat{O}=\hat{X}+\hat{Y}$, separating the environmental degrees of freedom which couple to the TLS into those fluctuating at frequencies below $\sim\,$MHz ($\hat{X}$) and those fluctuating at frequencies of the order of the energy splitting of the TLS, $E=\sqrt{\varepsilon^{2}+\Delta^{2}}~\approx 2\pi\cdot7\,$GHz ($\hat{Y}$).

In the eigenbasis of the TLS, Eq.~(\ref{eq:Hamiltonian1}) reads
\begin{align}
\label{eq:Hamiltonian2}\hat{\mathcal{H}}=&\,\frac{E}{2}\hat{\sigma}_{z}-\Omega^{0}_{\mathrm{R}}\cos\left(\omega_{d}t\right)\left(\hat{\sigma}_{z}\cos\theta-\hat{\sigma}_{x}\sin\theta\right)\nonumber\\
&+\frac{1}{2}\left(\hat{\sigma}_{z}\cos\theta-\hat{\sigma}_{x}\sin\theta\right)(\hat{X}+\hat{Y})+\hat{\mathcal{H}}_{\mathrm{b}}\, ,
\end{align}
where $\hat{\sigma}_{x}$ and $\hat{\sigma}_{z}$ are the Pauli matrices in the eigenbasis of the TLS, $\cos\theta=\varepsilon/E$ and $\sin\theta=\Delta/E$. All these are strain-dependent via $\varepsilon(\epsilon$), where $\epsilon$ is the strain at the position of the TLS. Taking into account the characteristic frequencies of $\hat{X}$ and $\hat{Y}$, the relevant terms are~\cite{HJ08}
\begin{align}
\label{eq:Hamiltonian3}\hat{\mathcal{H}}=&\,\frac{E}{2}\hat{\sigma}_{z}+\Omega_{\mathrm{R}}\cos\left(\omega_{d}t\right)\hat{\sigma}_{x}+\frac{1}{2}(\cos\theta\,\hat{\sigma}_{z}\hat{X}-\sin\theta\,\hat{\sigma}_{x}\hat{Y})\nonumber\\&+\hat{\mathcal{H}}_{\mathrm{b}}\, ,
\end{align}
where $\Omega_{\mathrm{R}}=\Omega^{0}_{\mathrm{R}}\sin\theta$ is the Rabi frequency. 

We are interested in the decay of Rabi oscillations, $\Gamma_{\mathrm{D}}$, which is the equivalent of $\Gamma_{2}$ in the rotating frame of reference. At resonance, $\omega_{d}=E$, the contribution of the high-frequency part $\hat{Y}$ to $\Gamma_{\mathrm{D}}$ is known to be $\frac{3}{4}\Gamma_{1}$,~\cite{HJ08} where $\Gamma_{1}=\frac{1}{2}\sin^{2}\theta\,S_{Y}(\omega=E)$ is the relaxation rate in the laboratory frame of reference, with $S_{Y}(\omega)$ being the spectral density of fluctuations in $\hat{Y}$. We now discuss the contribution of $\hat{X}$ to $\Gamma_{\mathrm{D}}$. To this end we consider the Hamiltonian 
\begin{align}
\label{eq:Hamiltonian4}\hat{\mathcal{H}}=\frac{E}{2}\hat{\sigma}_{z}+\Omega_{\mathrm{R}}\cos\left(\omega_{d}t\right)\hat{\sigma}_{x}+\frac{\cos\theta}{2}\hat{\sigma}_{z}\hat{X}+\hat{\mathcal{H}}_{\mathrm{b}}\, .
\end{align}
We now move to the rotating frame of reference by applying the unitary transformation $\hat{U}_{R}=e^{i\omega_{d}t\hat{\sigma}_{z}/2}$. Using the rotating wave approximation, in which counter-rotating terms with frequencies $2\omega_{d}$ are neglected, and assuming resonant driving ($\omega_{d}=E$), the transformed Hamiltonian
$\hat{\mathcal{H}}_{R}=\hat{U}_{R}\hat{\mathcal{H}}\hat{U}^{\dag}_{R}+i\frac{d\hat{U}_{R}}{dt}\hat{U}^{\dag}_{R}$ is
\begin{align}
\label{eq:Hamiltonian5}\hat{\mathcal{H}}_{R}=\frac{\Omega_{\mathrm{R}}}{2}\hat{\sigma}_{x}+\frac{\cos\theta}{2}\hat{\sigma}_{z}\hat{X}+\hat{\mathcal{H}}_{\mathrm{b}}\, .
\end{align}
We observe that in the rotating frame and at resonance, the noise is purely transverse. This transverse noise gives rise to relaxation in the rotating frame of reference,~\cite{HJ08} for which the golden rule yields
\begin{align}
\label{eq:Gamma_nu}\Gamma_{\nu}=\frac{1}{2}\cos^{2}\theta\,S_{X}(\omega=\Omega_{\mathrm{R}})\, ,
\end{align}
where $S_{X}(\omega)$ is the spectral density of fluctuations in $\hat{X}$. As usual, this results in a contribution of $\Gamma_{\nu}/2$ to $\Gamma_{\mathrm{D}}$. 

In an echo experiment, the same noise $\hat{X}$ is longitudinal and the dephasing rate is
\begin{align}
\label{eq:Echo dephasing rate1}\Gamma_{\mathrm{Echo}}=&\frac{1}{2}\cos^{2}\theta\,S_{X}(\omega\approx\Gamma_{\mathrm{Echo}})\, .
\end{align}
Equation~(\ref{eq:Echo dephasing rate1}) coincides with the golden rule result if $S_{X}(\omega)$ is flat on a scale of $\Gamma_{\mathrm{Echo}}$ around zero frequency. Otherwise, it provides self-consistently an order of magnitude estimate for the dephasing time. Both Eqs.~(\ref{eq:Gamma_nu}) and~(\ref{eq:Echo dephasing rate1}) are determined by the spectral density $S_{X}(\omega)$ at frequencies $\lesssim 1\,$MHz (see the experimental data below).

Based on the Ramsey dephasing observed and discussed in Refs.~\onlinecite{LJ16} and~\onlinecite{MS16}, $\hat{X}$ contains contributions from many slow fluctuators (thermal TLSs). Using standard estimations within the STM, one expects the relaxation rates $\Gamma_{1}$ of thermal TLSs (at $T=35\,$mK) to be smaller than $10\,$ms$^{-1}$.~\cite{LJ16,MS16} Consequently, these fluctuators are in the regime $\Gamma_{1}\ll\Omega_{\mathrm{R}}, \Gamma_{\mathrm{Echo}}$ and thus give rise to quasi-static noise, for which $\hat{X}$ is constant during each run of the experiment but fluctuates between different runs. These fluctuators will not contribute to $\Gamma_{\nu}$ and $\Gamma_{\mathrm{Echo}}$. However, due to their large number, their second-order contribution to pure dephasing may be important.

\begin{figure*}[htb!]
	\includegraphics[width=0.99\textwidth,height=8cm]{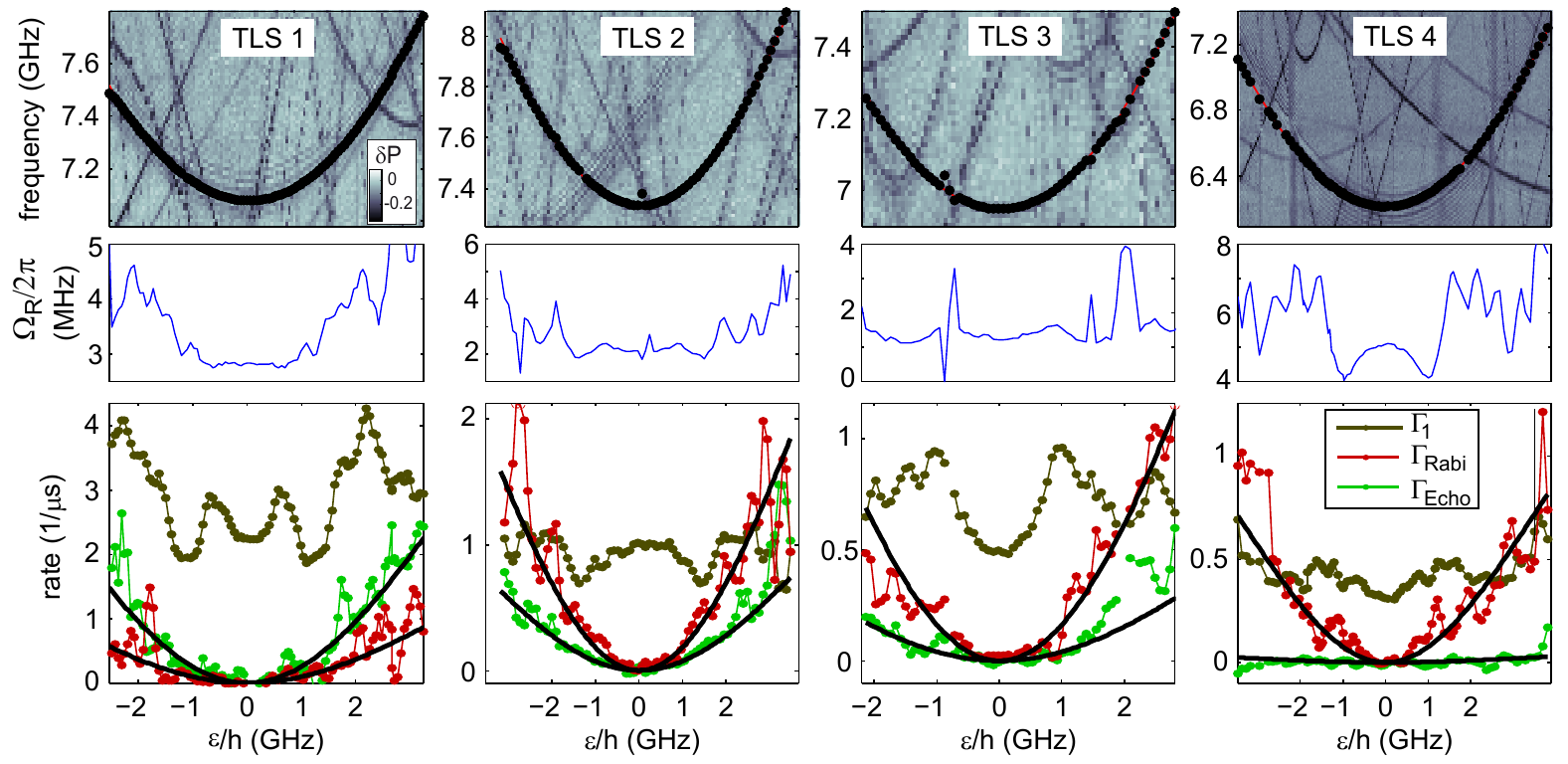} %[width=0.99\textwidth,height=10cm]
	\caption{Results obtained on four different TLSs. Top row: TLS swap-spectroscopy, showing the resonance frequencies of TLSs by a reduction $\delta P$ of the qubit population probability (dark traces in color-coded data). Superimposed dots are obtained from microwave spectroscopy, with the static TLS parameters resulting from hyperbolic fits. Middle row: Observed Rabi frequency $f_{\mathrm{R}}=\Omega_{\mathrm{R}}/2\pi$ for the probed TLSs driven at their resonance frequencies at fixed microwave power. Bottom row: TLS relaxation and pure dephasing rates as indicated in the legend, with quadratic fits to the Rabi oscillations and echo dephasing rates. The horizontal axes are calibrated to the asymmetry energy $\varepsilon/h$ of the probed TLS.}
	\label{fig1}
\end{figure*}

\begin{figure}[htb!]
	\includegraphics[width=0.99\columnwidth,height=9.5cm]{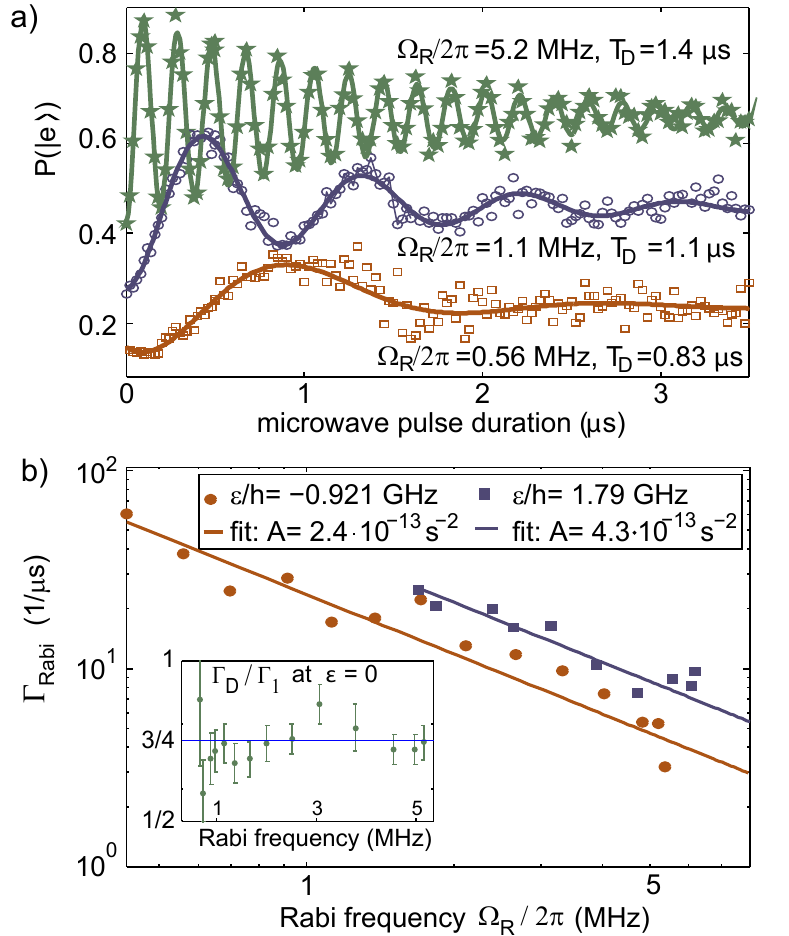} %[width=0.99\columnwidth,height=15cm]
	\caption{a) Rabi oscillations observed in TLS 2 at $\varepsilon/h=-0.921\,$GHz and at different drive powers (individual curves offset by 0.15 each for clarity). Rabi frequency $f_{\mathrm{R}}=\Omega_{\mathrm{R}}/2\pi$ and effective decay time $T_{\mathrm{D}}\equiv\Gamma^{-1}_{\mathrm{D}}$ quoted in the legend are extracted from fits to exponentially damped sinusoids (solid lines).
	b) Log-log plot of the dephasing rate $\Gamma_{\mathrm{Rabi}}=\Gamma_{\mathrm{D}}-\frac{3}{4}\Gamma_{1}$ as a function of the Rabi frequency. Solid lines are fits to $\Gamma_{\mathrm{Rabi}}=A/\Omega_{\mathrm{R}}$. Inset: At the symmetry point of the TLS, $\varepsilon=0$, the Rabi decay is only limited by energy relaxation such that $\Gamma_{\mathrm{D}}=\frac{3}{4}\Gamma_{1}$.}
	\label{fig2}
\end{figure} 

The noise due to these fluctuators can be treated classically, $\hat{X}\rightarrow X(t)=\sum_{j}v_{j}\alpha_{j}(t)$, where each fluctuator is represented by a random telegraph process (RTP) $\alpha_{j}(t)$ which randomly switches between the states $\alpha_{j}=\pm 1$ with flipping rate $\gamma_{1,j}$,~\cite{BJ09} and interacts with the probed TLS with a coupling strength $v_{j}$. As shown in Refs.~\onlinecite{LJ16} and~\onlinecite{MS16}, the interaction between the probed TLS and its closest thermal TLS, $v_{T}$, is of the order of a few MHz. As a result, close to the symmetry point $\varepsilon=0$ (thus $\cos\theta\ll 1$) one may assume $X\cos\theta\ll\Omega_{\mathrm{R}}$. We therefore expand
\begin{align}
\label{eq:energy splitting fluctuations}&\sqrt{\Omega^{2}_{\mathrm{R}}+X^{2}\cos^{2}\theta}\approx\,\Omega_{\mathrm{R}}+\frac{X^{2}\cos^{2}\theta}{2\Omega_{\mathrm{R}}}=\Omega_{\mathrm{R}}\nonumber\\&+\frac{\cos^{2}\theta}{2\Omega_{\mathrm{R}}}\sum_{j}v^{2}_{j}
+\frac{\cos^{2}\theta}{2\Omega_{\mathrm{R}}}\sum_{i\neq j}v_{i}v_{j}\alpha_{i}(t)\alpha_{j}(t)\, .
\end{align}
In contrast to the Ramsey dephasing,~\cite{LJ16,MS16} which is caused by individual thermal TLSs, the last term of Eq.~(\ref{eq:energy splitting fluctuations}) shows that the second-order contribution to the Rabi dephasing is due to pairs of thermal TLSs. However, since a product of two RTPs with flipping rates $\gamma_{1}$ and $\gamma_{2}$ is also a RTP with flipping rate $\gamma_{1}+\gamma_{2}$, the second-order contribution to the Rabi dephasing is essentially similar to the Ramsey dephasing.~\cite{LJ16,MS16} The decay law due to this low-frequency noise is
\begin{align}
\label{eq:Rabi decay1}f_{\mathrm{Rabi}}(t)=\frac{1}{2^{N}}\sum_{\{\xi_{k}\}}e^{it\frac{\cos^{2}\theta}{2\Omega_{\mathrm{R}}}\sum_{i\neq j}v_{i}v_{j}\xi_{i}\xi_{j}}\, ,
\end{align}
where $N$ is the number of thermal TLSs and the sum is over all the configurations of the variables $\xi_{k}=\pm1$. Similarly to the Ramsey dephasing, it is expected to be dominated by the few closest fluctuators, and the typical decay rate is~\cite{Comment1}
\begin{align}
\label{eq:pure dephasing rate}\Gamma^{(2)}_{\varphi}\approx\frac{v^{2}_{T}}{\Omega_{\mathrm{R}}}\cos^{2}\theta\, .
\end{align}

The Rabi decay rate is given by the sum of the three contributions discussed above,
\begin{align}
\label{eq:Rabi decay rate1}\Gamma_{\mathrm{D}}=\frac{3\Gamma_{1}}{4}+\frac{\Gamma_{\nu}}{2}+\Gamma^{(2)}_{\varphi}\, .
\end{align}
We now discuss the experimental results for this decay rate. The distinct strain dependence of the first and the other two terms of Eq.~(\ref{eq:Rabi decay rate1}) allows us to separate the effect of the noise at the two spectral ranges discussed above (i.e., below $\sim\,$MHz and at GHz frequencies).

The TLSs studied here are contained in the amorphous AlO$_x$ tunnel barrier of a Josephson junction, which is part of a superconducting phase qubit circuit.~\cite{Steffen06} We apply mechanical strain to the qubit chip by means of a piezo actuator,~\cite{GGJ12} allowing us to tune the asymmetry energy $\varepsilon$ of the TLSs while their resonance frequencies are tracked with the qubit,~\cite{LisenfeldNatureComm} as shown in the top row of Fig.~\ref{fig1}. At each strain value, standard microwave pulse sequences are applied~\cite{LJ16} to measure the energy relaxation rate $\Gamma_{1}$ and the dephasing rates of Rabi oscillations and echo signals, $\Gamma_{\mathrm{Rabi}}$ and $\Gamma_{\mathrm{Echo}}$, of the probed TLS (bottom row of Fig.~\ref{fig1}). In addition, we present the frequency of Rabi oscillations in the middle row of Fig.~\ref{fig1} for a fixed microwave driving power. Peaks and dips, which appear symmetrically in these data with respect to $\varepsilon=0$, mainly originate from the frequency dependence of the transmitted microwave power due to cable resonances. When these fluctuations are smoothened out, the Rabi frequency tends to increase with $\varepsilon$. This is because the TLS is driven via a transition that involves a virtual state of the qubit, such that the effective driving strength depends on the detuning between the TLS and the qubit,~\cite{LJ10b} where the latter was tuned to the fixed frequency of $8.8\,$GHz during all measurements. This increase is partly compensated by the factor $\sin\theta=\Delta/E$ appearing in the definition of the Rabi frequency.

The strain dependence of the energy relaxation and echo dephasing rates was discussed in Ref.~\onlinecite{LJ16}. Here we focus on the Rabi dephasing rate. Due to limitations in the applicable microwave power, the Rabi frequency is at maximum of order $5\,$MHz, such that only a few oscillations can be observed during the coherence time of the TLS [see Fig.~\ref{fig2}(a)]. Together with the experimental measurement uncertainties it therefore becomes practically impossible to determine the exact functional form of the decay envelope. We thus extract an effective decay rate $\Gamma_{\mathrm{D}}$ from a fit to an exponentially damped sinusoid.

At the symmetry point $\varepsilon=0$ ($\cos\theta=0$), one expects $\Gamma_{\mathrm{D}}=\frac{3}{4}\Gamma_{1}$, so that the Rabi decay rate is independent of the Rabi frequency. The experimental observation is in good agreement with this prediction, as shown in the inset of Fig.~\ref{fig2}(b). To analyze the effect of the noise at frequencies below the Rabi frequency, one has to subtract this contribution of fluctuations at GHz frequencies from $\Gamma_{\mathrm{D}}$. In the bottom row of Fig.~\ref{fig1} we show the Rabi dephasing rate $\Gamma_{\mathrm{Rabi}}\equiv\Gamma_{\mathrm{D}}-\frac{3}{4}\Gamma_{1}$. For all investigated TLSs, $\Gamma_{\mathrm{Rabi}}$ vanishes at the symmetry point and increases quadratically with $\varepsilon$. This quadratic strain dependence is also predicted by the above model, which shows that both $\Gamma_{\nu}$ and $\Gamma^{(2)}_{\varphi}$ scale as $\cos^{2}\theta=(\varepsilon/E)^2\approx (\varepsilon/\Delta)^2$ [Eqs.~(\ref{eq:Gamma_nu}) and~(\ref{eq:pure dephasing rate})]. Table~\ref{table1} summarizes the measured tunneling energies $\Delta$, the deformation potentials $\partial\varepsilon/\partial V$ where $V$ is the applied piezo voltage, and the relaxation times $T_{1}$ measured at $\varepsilon=0$. The strength of Rabi and echo dephasing rates is obtained from quadratic fits to $\Gamma_{i}= A_{i}(\varepsilon/E)^{2}$, where $i=\{\mathrm{Rabi},\mathrm{Echo}\}$. 

\begin{table} [b!]
	\footnotesize 
	\begin{tabular}{|c|c|c|c|c|c|c|} \hline \  
		TLS & $\Delta/h$  & $(\partial \varepsilon / \partial V)/h$   & $T_1$ @ $\varepsilon=0$ & $A_\mathrm{Rabi}$ & $A_\mathrm{Echo}$ & $\frac{A_\mathrm{Rabi}}{A_\mathrm{Echo}}$ \\ 
		&  (GHZ)  & (MHz/V) & ($\mu s)$ &(MHz)&(MHz)&  \\ \hline
		1 &  7.075(3) & 115.5(0.7) & 0.44 &  5 &  13 & 0.38\\
		2 &  7.335(7) & 180.3(2) & 0.99 & 10 &   4 & 2.5\\
		3 &  6.947(5) & 156.7(2) & 2 &  8 &   2 &  4\\
		4 &  6.217(3) & 146.8(0.6) & 3.2 &  3 & $<$0.1 & $>$30\\
		\hline \end{tabular} \normalsize \caption{Parameters of the four TLSs investigated here.} \label{table1}
\end{table}

We now compare the values extracted for $A_{\mathrm{Rabi}}$ and $A_{\mathrm{Echo}}$ with the expressions derived from the above model, 
\begin{align}
\label{eq:Rabi dephasing rate2}&A_{\mathrm{Rabi}}=\frac{\Gamma_{\mathrm{Rabi}}}{\cos^{2}\theta}=\frac{\Gamma_{\nu}/2+\Gamma^{(2)}_{\varphi}}{\cos^{2}\theta}=\frac{S_{X}(\Omega_{\mathrm{R}})}{4}+\frac{v^{2}_{T}}{\Omega_{\mathrm{R}}}\, ,\\
\label{eq:Echo dephasing}&A_{\mathrm{Echo}}=\frac{\Gamma_{\mathrm{Echo}}}{\cos^{2}\theta}=\frac{S_{X}(\Gamma_{\mathrm{Echo}})}{2}\, .
\end{align}
Further information on $A_{\mathrm{Rabi}}$ comes from the inverse dependence of the Rabi dephasing rate on the Rabi frequency, $\Gamma_{\mathrm{Rabi}}=A/\Omega_{\mathrm{R}}$, observed at fixed $\varepsilon$ and displayed in Fig.~\ref{fig2}(b) for TLS 2, with fit parameters $A$ listed in the legend. Comparing this observation with Eq.~(\ref{eq:Rabi dephasing rate2}), we identify two possibilities: I) The first term of Eq.~(\ref{eq:Rabi dephasing rate2}) dominates, in which case $S_{X}(\Omega_{\mathrm{R}})\propto\Omega^{-1}_{\mathrm{R}}$ ($1/f$ noise). II) The second contribution dominates. In the first case $\Gamma_{\mathrm{Echo}}\gg\Gamma_{\mathrm{Rabi}}$ because $\Gamma_{\mathrm{Echo}}\ll\Omega_{\mathrm{Rabi}}$, which is inconsistent with the experimental observations. Thus we adopt the second scenario. Indeed, with $v_{T}$ estimated by several MHz,~\cite{LJ16,MS16} the contribution $v^{2}_{T}/\Omega_{\mathrm{R}}$ due to thermal standard TLSs gives the correct order of magnitude for $A_{\mathrm{Rabi}}$, and therefore this scenario seems plausible. Moreover, according to our previous results,\cite{LJ16} the strain dependence and the ratio between the Ramsey and the echo dephasing rates are inconsistent with a $1/f$ spectrum at frequencies around $\Gamma_{\mathrm{Echo}}$, but rather suggest a flat spectrum. It is therefore improbable that $S_{X}(\Omega_{\mathrm{R}})\propto\Omega^{-1}_{\mathrm{R}}$, leading to the conclusion that $S_{X}(\Omega_{\mathrm{R}})\ll v^{2}_{T}/\Omega_{\mathrm{R}}$.   

What could then be learned from comparing $A_{\mathrm{Rabi}}$ with $A_{\mathrm{Echo}}$? Based on our previous results in Ref.~\onlinecite{LJ16}, we try to associate the two contributions to $\Gamma_{\mathrm{Rabi}}$ with the two types of TLSs discussed in Refs.~\onlinecite{MS16,SM13}. In Ref.~\onlinecite{LJ16,MS16} we attributed the Ramsey dephasing to an ensemble of slow (quasi-static) thermal TLSs (denoted as $\tau$-TLSs), with parameters consistent with the STM, for which the echo protocol should be more efficient than observed. This discrepancy was explained by the existence of a few fluctuators, with maximum relaxation rates of the order of $10\,(\mu$s)$^{-1}$ (denoted as $S$-TLSs). In contrast to the standard fluctuators, these fast fluctuators contribute to $S_{X}(\omega)$ at MHz frequencies. Let us assume the existence of a single fast fluctuator with relaxation rate $\gamma_{1}$ and coupling constant to the probed TLS $v_{S\tau}$, for which $S_{X}(\omega)=2v^{2}_{S\tau}\gamma_{1}/(\omega^{2}+\gamma^{2}_{1})$. If the second contribution to $A_{\mathrm{Rabi}}$ is dominant, the observation that $A_{\mathrm{Rabi}}$ and $A_{\mathrm{Echo}}$ are comparable (see Fig.~\ref{fig1} and table~\ref{table1}) implies that $S_{X}(\Gamma_{\mathrm{Echo}})\gg S_{X}(\Omega_{\mathrm{R}})$. Since $\Omega_{\mathrm{R}}\gg\Gamma_{\mathrm{Echo}}$, one obtains the condition $\Omega_{\mathrm{R}}>\gamma_{1}$, which sets an upper bound of $\sim 10\,(\mu$s)$^{-1}$ for $\gamma_{1}$ of the fast relaxing fluctuator.

In summary, we have measured the decay of Rabi oscillations as a function of the asymmetry energy of individual TLSs, which is controlled by applying an external strain. We employ a theoretical model based on interacting TLSs and find agreement with the experimentally observed magnitude of the Rabi dephasing rate and its dependence on the applied strain and on the Rabi frequency. In conjunction with measurements of energy relaxation, Ramsey and echo dephasing,~\cite{LJ16,MS16} Rabi noise spectroscopy provides information about the spectrum of the environment to which TLSs couple, within three different spectral ranges. This allows one to distinguish between contributions from distinct environmental degrees of freedom. Such information is important for minimizing noise due to TLSs in various nano-devices, for exploiting TLSs as useful degrees of freedom, and for a basic understanding of amorphous systems at low temperatures.    

This work was supported by the German-Israeli Foundation (GIF), grant 1183/2011, by the Israel Science Foundation (ISF), grant 821/14, and by the Deutsche Forschungsgemeinschaft (DFG), grants SH 81/2-1 and LI2446/1-1. AB acknowledges support from the Helmholtz International Research School for Teratronics (HIRST) and the Landesgraduiertenf\"orderung-Karlsruhe (LGF).


\begin{references}
\bibitem{Comment1} Here we use the term "rate" to refer to the inverse of the time for which the signal has decayed to $e^{-1}$. The decay in this case is not exponential but rather Gaussian.
\end{references}
\end{document}